\documentstyle[11pt,deutsch]{article}

\captionsenglish
\begin{document}
\begin{center}
%\vspace{10mm}
%{\Large Flavour Distributions in the Nucleons:}
{\Large Flavour distributions in the nucleons:}

%{\Large SU(2) Sea Asymmetry or Isospin Symmetry Breaking?}
{\Large SU(2) sea asymmetry or isospin symmetry breaking?}
\vspace{10mm}
\renewcommand{\thefootnote}{\fnsymbol{footnote}}
{\normalsize Bo-Qiang Ma\footnote{
Fellow of Alexander von Humboldt foundation,
on leave from
Institute of High Energy Physics, Academia Sinica, P.O.Box 918(4),
Beijing 100039, China},
Andreas Sch\"afer, and Walter Greiner}

\vspace{8mm}
{\large
        Institut f\"ur Theoretische Physik der
        Universit\"at Frankfurt am Main, Postfach 11 19 32,
        D-6000 Frankfurt, Germany}
\vspace{8mm}

\end{center}
%\renewcommand{\baselinestretch}{1.0}

%\break
%\begin{minipage}[t] {105mm}
%{\large \bf Abstract } \\
{\large \bf Abstract }
The Gottfried sum
rule violation reported by the New
Muon Collaboration(NMC) was interpreted as indication for a flavour
asymmetry of the sea quark in the nucleons. We investigate the
alternative possibility that isospin
symmetry between the proton and the neutron is breaking.
We examine systematically the consequences of this possibility
for several processes, namely, neutrino deep inelastic
scattering,
charged pion Drell-Yan process, proton Drell-Yan process, and
semi-inclusive
deep inelastic
scattering, and conclude that a decision between the two alternative
explanations is possible.
%\end{minipage}

\noindent
To be published in Phys.Rev.D

\break
\renewcommand{\theequation}{\thesection.\arabic{equation}}

\renewcommand{\thesection}{\Roman{section}.}
\section{INTRODUCTION}
\renewcommand{\thesection}{\arabic{section}}

Recently, the measurement of the Gottfried sum \cite{Got67}
by the New Muon Collaboration (NMC) \cite{NMC91}
has inspired a number of investigations on the sea-quark
flavour distributions
in of
nucleons\cite{Pre91}-\cite{Kum92}.
%\cite{Pre91,Kum91a,Hen90,Sig91,Ste91,Mar90,Ans91}
%\cite{Ma92,Kap91,Epe92}.
The Gottfried sum is defined, in terms
of the proton and neutron structure functions $F_{2}(x)$, as
$S_{G}=\int_{0}^{1}(F_{2}^{p}(x)-F_{2}^{n}(x))\;dx/x$, which,
when expressed in terms of quark momentum distributions
$q_{i}^{N}(x)$ in the nucleon N, reads
\begin{equation}
S_{G}=\int_{0}^{1}\sum_{i}e_{i}^{2}[q_{i}^{p}(x)+\overline{q}_{i}^{p}(x)
-q_{i}^{n}(x)-\overline{q}_{i}^{n}(x)]\;dx,
\end{equation}
where $e_{i}$ is the charge of a quark of flavour i. Using flavour
conservation, isospin symmetry between
the proton and the neutron and further the flavour
distribution symmetry in the sea, one arrives at the Gottfried sum
rule (GSR), $S_{G}=1/3$. In the NMC experiment, the value of $S_{G}$ was
determined from $F_{2}^{p}-F_{2}^{n}$ expressed as
\begin{equation}
F_{2}^{p}-F_{2}^{n}=2F_{2}^{D}(1-F_{2}^{n}/F_{2}^{p})/
(1+F_{2}^{n}/F_{2}^{p}),
\end{equation}
where the ratio $F_{2}^{n}/F_{2}^{p}=2F_{2}^{D}/F_{2}^{p}-1$ was
determined from the deuteron/proton cross-section ratio measured
in the experiment. The data cover the kinematic range of
$x=0.004-0.8$ for $Q^{2}=4$GeV$^{2}$.  Assuming a smooth
extrapolation of the data $F_{2}^{n}/F_{2}^{p}$ from $x=0.8$ to $x=1$,
and adopting a Regge behavior $ax^{b}$ for
$F_{2}^{p}-F_{2}^{n}$, a flavour nonsinglet, in the region
$x=0.004-0.15$ and then extrapolating it to $x=0$, the NMC reported
the value for the Gottfried sum
\begin{equation}
S_{G}=0.240\pm0.016,
\end{equation}
which is significantly below the simple quark-parton-model result of
1/3.

Several different effects have been proposed as the source for this
discrepancy between the NMC data and the GSR.
%ottfried sum rule.
In several early
works \cite{Pre91,Kum91a,Hen90,Sig91} the NMC
data were claimed as evidence for a flavour distribution
asymmetry in the sea of nucleons, i.e., an excess of d\=d
over u\=u pairs in the proton\cite{Pre91,Kum91a,Hen90,Sig91}.
This suppression of $u\overline{u}$ pairs could be due
to Pauli exclusive principle and
%may contribute to this sea flavour
%asymmetry from more
%suppression of $u\overline{u}$ pairs than $d\overline{d}$ pairs
%due to
the excess of
a u valence quark in the proton\cite{Fie77,Mel91}.
A flavour SU(2) asymmetry in the sea can also be attributed to the
pionic contribution as an excess of $p \rightarrow n+\pi^{+}$ over
$p \rightarrow \Delta^{++}+\pi^{-}$ (or $\pi^{+}$ over
$\pi^{-}$)\cite{Kum91a,Hen90,Sig91,Ste91,Hwa91},
or from a more microscopic point of view to the excess of
$u\rightarrow d+\pi^{+}$ over $u \rightarrow u+\pi^{0}$
and $d\rightarrow u+\pi^{-}$ over
$d\rightarrow d+\pi^{0}$\cite{Eic92,Wak91}.
Other possibilities are that the smaller NMC result for the GSR
is due to the
unjustified $x \rightarrow 0$ extrapolation of the
available data \cite{Mar90}
%, or explained as
or due to a small admixture of vector
diquarks in a particular quark-diquark model of the nucleon\cite{Ans91}.
%or understood from a meson exchange picture in generating the
%strangeness content
%in protons\cite{Hwa91}.
%These two explanations are not in the conventional quark-parton language
%and we will not pay much attention in this paper.
More recently, it has been observed \cite{Ma92} that the violation of the
GSR
%Gottfried sum rule
could also
be due to isospin symmetry breaking between the proton
and neutron while the sea flavour distribution symmetry
of nucleons is preserved.
According to this explanation there are more sea quarks in the
neutron than in the proton. Nuclear effects have also been examined
\cite{Kap91,Epe92,Zol92,Kum91b} and it has been
suggested that,
%the nuclear effects\cite{Epe92},
e.g., the mesonic exchanges in deuteron \cite{Kap91}
could account for the difference between the NMC data and the
GSR,
%Gottfried sum rule,
although
nuclear shadowing
corrections
may lead to more smaller values of $S_{G}$
than reported by NMC\cite{Zol92,Kum91b}.

In this paper we discuss the possibility of explicit isospin
symmetry breaking between the proton and neutron,
i.e., $u^{p}(x)\neq d^{n}(x)$ etc..
For simplicity we shall always assume sea flavour symmetry
($\overline{u}^{p}(x)=\overline{d}^{p}(x)$) although both effects could
occur simultaneously. We shall review the various experiments
proposed to determine the origin of the small NMC data and
analyse what results would be expected for our assumptions.
%The possibilities to check some of the above explanations
%through other reactions have been discussed by several
%groups\cite{Kum91b,Ell91,Lev91,Ma92}.
%\noindent
%[1].
In the paper of Kumano and Londergan \cite{Kum91b}
it was argued that  a combination
of neutrino structure functions of proton and deuteron,
or an expression of the relative Drell-Yan cross sections
%\begin{equation}
%\frac{1}{2}[F_{2}^{\nu D}(x)-xF_{3}^{\nu D}(x)]-
%[F_{2}^{\nu P}(x)-xF_{3}^{\nu P}(x)]
%=2x[\overline{d}(x)-\overline{u}(x)],
%\end{equation}
%was proposed to measure the excess of d\=d over u\=u.
%\noindent [2].
%Also in the paper the quantity $\overline{u}(x)-\overline{d}(x)$
%was suggested to be measured
for
charged pions scattering from nuclear targets,
%\begin{equation}
%R_{sea}=
%\frac{4[\tilde{\sigma}(\pi^{+}A_{1})-\tilde{\sigma}(\pi^{+}A_{0})]
%+\tilde{\sigma}(\pi^{-}A_{1})-\tilde{\sigma}(\pi^{-}A_{0})}
%{\tilde{\sigma}(\pi^{+}A_{0})-\tilde{\sigma}(\pi^{-}A_{0})}
%=\frac{10\varepsilon_{1}(\overline{u}-\overline{d})}
%{u_{V}+d_{V}}.
%\end{equation}
%\noindent [3].
are sensitive to the quantity $\overline{d}-\overline{u}$.
Ellis and Stirling \cite{Ell91} suggested
to compare
%a new Drell-Yan experiment comparing
pp$\rightarrow$\cal l$^{+}$\cal l$^{-}$X and
pn$\rightarrow$\cal l$^{+}$\cal l$^{-}$X  to distinguish between
the SU(2) asymmetry in the sea and a non-Regge behavior
at small $x$.
% values.
%\noindent [4].
Levelt, Mulders and Schreiber \cite{Lev91} discussed the
possible measurement of the asymmetry in the sea quark distribution
using semi-inclusive leptoproduction of charged pions and,
more general, charged
hadrons, to check for isospin asymmetries.
%y explanation from semi-inclusive data.
Those above processes are discussed before the
observation \cite{Ma92} that
isospin symmetry breaking between the proton and the neutron could be
an alternative source for the
%Gottfried sum rule
GSR
violation.
The purpose of this paper is to examine the influences
from p-n isospin symmetry breaking
for above processes.
It is
necessary to point out that the nuclear mesonic exchange
explanation \cite{Kap91}
suggested also an excess of mesonic contributions
% besides from
over the
%SU(2)
isospin symmetric p-n in deuteron, and thus
should give the same results
as isospin symmetry breaking if free
neutrons
are not involved in the measurements.

\renewcommand{\thesection}{\Roman{section}.}
\section{DEEP INELASTIC NEUTRINO SCATTERING}
%\section{Deep inelastic neutrino scattering}
\renewcommand{\thesection}{\arabic{section}}

The possibility to distinguish between a SU(2) sea asymmetry
and p-n isospin symmetry breaking
through deep inelastic neutrino and anti-neutrino scattering
on protons and deuterons
has been discussed in Ref.\cite{Ma92}. If the SU(2) sea is asymmetry
%explanation the source for the Gottfried sum rule
the violation of GSR indicates
%is due to
an excess of $d\overline{d}$ over $u\overline{u}$;
%in the sea of proton;
i.e.,
\begin{equation}
\int_{0}^{1}[\overline{u}(x)-\overline{d}(x)]
=-0.140\pm0.024.
\label{eq:su2sea}
\end{equation}
Whereas for p-n isospin symmetry breaking the
violation of GSR is due to an excess of sea quarks in
neutrons over those in protons while preserving the SU(2) symmetry
in the sea of nucleons; i.e.,
\begin{equation}
\int_{0}^{1}[\overline{q}^{p}(x)-\overline{q}^{n}(x)]
=-0.084\pm0.014.
\label{eq:pnsea}
\end{equation}
Both explanations are fitted to the observed
%, though give a same value of the Gottfried sum
$S_{G}$ but
may give very different values for some linear combinations of
neutrino structure functions from protons and deuterons. In
Ref.\cite{Ma92} a new sum, defined by
\begin{equation}
S=\int_{0}^{1}{[(F_{2}^{\nu p}+F_{2}^{\overline{\nu}p})-
\frac{1}{2}(F_{2}^{\nu D}+F_{2}^{\overline{\nu} D})]}dx,
\label{eq:ns}
\end{equation}
was suggested. This new sum is zero for an asymmetric sea
explanation and
 $4\int_{0}^{1}[\overline{q}^{p}(x)-\overline{q}^{n}(x)]=
-0.336\pm0.058$ for p-n symmetry breaking.
Some other combinations, such as those indicated
in
Ref.\cite{Ma92}, may also
be chosen
%as a new sum
to distinguish between the two explanations.

The linear  combination
of neutrino structure functions from protons and deuterons,
\begin{equation}
\frac{1}{2}[F_{2}^{\nu D}(x)-xF_{3}^{\nu D}(x)]-
[F_{2}^{\nu P}(x)-xF_{3}^{\nu P}(x)],   \label{eq:KL}
\end{equation}
which was proposed by Kumano and Londergan \cite{Kum91b} to
measure
the difference $\overline{u}-\overline{d}$,
should be
\begin{equation}
2x[\overline{d}(x)-\overline{u}(x)]
\end{equation}
in the case of SU(2) asymmetry in the sea. This combination,
% Eq.\ref{eq:KL},
when expressed
in terms of quark momentum distributions, should read
\begin{equation}
2x[\overline{u}^{n}(x)-\overline{u}^{p}(x)].
\end{equation}
While it gives
\begin{equation}
2x[\overline{q}^{n}(x)-\overline{q}^{p}(x)]
\end{equation}
in the case of isospin symmetry breaking between the proton
and the neutron. The two cases give different values for
Eq.(\ref{eq:KL}). From Eq.(\ref{eq:su2sea}) and Eq.(\ref{eq:pnsea}),
one can see that
the prediction for Eq.(\ref{eq:KL})
 is less for
n-p isospin symmetry than for SU(2) sea asymmetry.
% for the expression Eq.(\ref{eq:KL}).

\renewcommand{\thesection}{\Roman{section}.}
\section{DRELL-YAN PROCESSES}
%\section{Drell-Yan process by charged pion from nuclear targets}
\renewcommand{\thesection}{\arabic{section}}
\setcounter{equation}{0}

%\renewcommand{\thesection}{\Roman{section}.}
%\subsection{DRELL-YAN PROCESS BY CHARGED PION FROM NUCLEAR TARGETS}
\subsection{Drell-Yan processes by charged pion from nuclear targets}
%\renewcommand{\thesection}{\arabic{section}}
%\setcounter{equation}{0}

%Also in the paper the quantity $\overline{d}(x)-\overline{u}(x)$
%was suggested to be measured
%by Drell-Yan processes initiated by
%charged pion scattering from nuclear targets,
%\begin{equation}
%R_{sea}=
%\frac{4[\tilde{\sigma}(\pi^{+}A_{1})-\tilde{\sigma}(\pi^{+}A_{0})]
%+\tilde{\sigma}(\pi^{-}A_{1})-\tilde{\sigma}(\pi^{-}A_{0})}
%{\tilde{\sigma}(\pi^{+}A_{0})-\tilde{\sigma}(\pi^{-}A_{0})}
%=\frac{10\varepsilon_{1}(\overline{u}-\overline{d})}
%{u_{V}+d_{V}}.
%\end{equation}
%\noindent [3].
Also in Ref.\cite{Kum91b}
the quantity $\overline{d}(x)-\overline{u}(x)$
was suggested to be measured
by Drell-Yan processes
$\pi^{\pm}$A$\rightarrow$\cal l$^{+}$\cal l$^{-}$X
initiated by
charged pion scattering from nuclear targets. For large
$x_{\pi}$, say $x_{\pi}>0.4$,
%the contributions from the sea quark distributions
%in the pion can be negligible, and in this region
the cross section per nucleon for $\pi^{+}$ and $\pi^{-}$  scattering
from nuclear targets will be proportional to
\begin{equation}
\begin{array}{clcr}

\tilde{\sigma}(\pi^{+}A)=V_{\pi}[\frac{1}{2}
(d^{n}+4\overline{u}^{n}+
d^{p}+4\overline{u}^{p})
+\varepsilon
(d^{n}+4\overline{u}^{n}-
d^{p}-4\overline{u}^{p});\\

\tilde{\sigma}(\pi^{-}A)=V_{\pi}[\frac{1}{2}
(4u^{n}+\overline{d}^{n}+
4u^{p}+\overline{d}^{p})
+\varepsilon
(4u^{n}+\overline{d}^{n}-
4u^{p}-\overline{d}^{p}),
\end{array}
\end{equation}
where $\varepsilon\equiv N/A-1/2$ is the neutron excess of the target
with $A$ nucleon and $V_{\pi}$ is the valence distribution of the
pion. Thus one can define the ratio
\begin{equation}
R_{sea}=
\frac{4[\tilde{\sigma}(\pi^{+}A_{1})-\tilde{\sigma}(\pi^{+}A_{0})]
+\tilde{\sigma}(\pi^{-}A_{1})-\tilde{\sigma}(\pi^{-}A_{0})}
{\tilde{\sigma}(\pi^{+}A_{0})-\tilde{\sigma}(\pi^{-}A_{0})}.
\label{eq:pidy}
\end{equation}
In the case of an asymmetric SU(2) sea, one assumes the p-n isospin
symmetry, i.e.,
\begin{equation}
\begin{array}{clcr}
u=u^{p}\leftrightarrow d^{n};\;\;\;d=d^{p}\leftrightarrow u^{n};\\
\overline{u}=\overline{u}^{p}\leftrightarrow \overline{d}^{n};
\;\;\;\overline{d}=\overline{d}^{p}\leftrightarrow \overline{u}^{n},
\label{eq:pnis}
\end{array}
\end{equation}
then one gets
\begin{equation}
R_{sea}=
\frac{10(\varepsilon_{1}-\varepsilon_{0})
(\overline{u}-\overline{d})}
{u_{V}+d_{V}}.      \label{eq:fsea}
\end{equation}
In the case of p-n isospin symmetry breaking, we assume that only
the valence quarks  preserve the isospin symmetry between the proton
and the neutron while sea quarks do not, then
it can be found
\begin{equation}
R_{sea}=
\frac{50(\varepsilon_{1}-\varepsilon_{0})
(\overline{q}^{p}-\overline{q}^{n})}
{3(u_{V}+d_{V})}.  \label{eq:iso}
\end{equation}
{}From eqs.\ref{eq:su2sea}  and \ref{eq:pnsea} we know the relation
\begin{equation}
\int_{0}^{1}[\overline{q}^{p}(x)-\overline{q}^{n}(x)]
=\frac{3}{5}\int_{0}^{1}[\overline{u}(x)-\overline{d}(x)]
\label{eq:relation}
\end{equation}
between p-n isospin symmetry breaking and an asymmetric SU(2) sea,
thus we can reasonable assume
\begin{equation}
[\overline{q}^{p}(x)-\overline{q}^{n}(x)]
=\frac{3}{5}[\overline{u}(x)-\overline{d}(x)].
\label{eq:relation2}
\end{equation}
Substituting this
relation into Eq.(\ref{eq:iso}) we arrive at Eq.(\ref{eq:fsea}). Thus
the two explanations give approximately the same result for the
expression Eq.(\ref{eq:pidy}) introduced by Kumano and Londergan.

%\renewcommand{\thesection}{\Roman{section}.}
%\subsection{DRELL-YAN PROCESS BY PROTON SCATTERING}
\subsection{Drell-Yan processes by proton scattering}

We now turn our attention to the Drell-Yan processes for protons
scattered from proton and deuteron targets, which are discussed by
Ellis and Stirling\cite{Ell91}, and also by Kumano and
Londergan \cite{Kum92} recently. The cross section of the process
%$pN\rightarrow \cal l^{+}\cal l^{-} X$
pN$\rightarrow$\cal l$^{+}$\cal l$^{-}$X
is sensitive to the sea antiquark
distributions:
\begin{equation}
s\frac{d^{2}\sigma^{pN}}{d\sqrt{\tau}dy}=
\frac{8\pi\alpha^{2}}{9\sqrt{\tau}}\sum_{i}e_{i}^{2}{q_{i}(x_{1},M)
\overline{q}_{i}(x_{2},M)+[1\leftrightarrow 2]},
\end{equation}
where $\tau=M^{2}/s$, $x_{1}=\sqrt{\tau}e^{y}$ and
$x_{2}=\sqrt{\tau}e^{-y}$. We ignore, following Ellis-Stirling and
Kumano-Londergan, the contributions from strange and charm quarks and
retain only the dominant "'valence-sea'' annihilation terms in the
remainder, then we get, at $y=0$,
\begin{equation}
\begin{array}{clcr}

\sigma^{pp}\equiv s\frac{d^{2}\sigma^{pp}}{d\sqrt{\tau}dy}=
\frac{8\pi\alpha^{2}}{9\sqrt{\tau}}
(\frac{8}{9}u_{V}^{p}\overline{u}^{p}+
 \frac{2}{9}d_{V}^{p}\overline{d}^{p});\\

\sigma^{pn}\equiv s\frac{d^{2}\sigma^{pn}}{d\sqrt{\tau}dy}=
\frac{8\pi\alpha^{2}}{9\sqrt{\tau}}
[\frac{4}{9}(u_{V}^{p}\overline{u}^{n}+u_{V}^{n}\overline{u}^{p})+
 \frac{1}{9}(d_{V}^{p}\overline{d}^{n}+d_{V}^{n}\overline{d}^{p})].
\end{array}
\end{equation}
The p-n cross-section asymmetry
\begin{equation}
A_{DY}=\frac{\sigma^{pp}-\sigma^{pn}}{\sigma^{pp}+\sigma^{pn}}
\label{eq:ady}
\end{equation}
discussed by Ellis-Stirling becomes
\begin{equation}
A_{DY}=\frac{(4u_{V}-d_{V})(\overline{u}-\overline{d})+
(u_{V}-d_{V})(4\overline{u}-\overline{d})}
{(4u_{V}+d_{V})(\overline{u}+\overline{d})+
(u_{V}+d_{V})(4\overline{u}+\overline{d})}    \label{eq:adysu2}
\end{equation}
for a SU(2) asymmetric sea. It was found by Ellis and
Stirling that the quantity  Eq.(\ref{eq:adysu2}) can change sign
for an asymmetric sea whereas it is positive in the
case that the violation of GSR
%Gottfried sum rule violation
is due to the unjustified
$x\rightarrow 0$ extrapolation of the data.
We assume instead that only
the valence quarks  preserve the isospin symmetry between the proton
and the neutron while sea quarks do not. Thus we find
\begin{equation}
A_{DY}=
\frac{(4u_{V}-d_{V})5(\overline{q}^{p}-\overline{q}^{n})/3
+(u_{V}-d_{V})(\overline{q}^{p}+8\overline{q}^{n})/3}
{9(\sigma^{pp}+\sigma^{pn})}.  \label{eq:adypn}
\end{equation}
Comparing Eq.(\ref{eq:adypn}) with Eq.(\ref{eq:adysu2}), one can find that
the p-n cross section asymmetry could also change sign
in the case of isospin symmetry breaking between the proton and the
neutron. Using the relation Eq.(\ref{eq:relation2}) and
adopting
$(4\overline{u}-\overline{d})\sim (\overline{q}^{p}+8\overline{q}^{n})/3$,
we see
that Eq.(\ref{eq:adypn}) will give approximately the same value as
Eq.(\ref{eq:adysu2}). Therefore the Drell-Yan processes discussed
in the literature \cite{Kum91b,Ell91,Kum92} are not sufficient
to distinguish between
an asymmetric SU(2) sea and p-n
isospin symmetry breaking for charged pion scattering
Eq.(\ref{eq:pidy}) as well as for
%both the quantities Eq.(\ref{eq:pidy}) in charged pion scattering
%and Eq.(\ref{eq:ady}) in
proton scattering Eq.(\ref{eq:ady}).

\renewcommand{\thesection}{\Roman{section}.}
%\section{Semi-inclusive deep inelastic scattering}
\section{SEMI-INCLUSIVE DEEP INELASTIC SCATTERING}
\renewcommand{\thesection}{\arabic{section}}
\setcounter{equation}{0}

It has been suggested by Levelt, Mulders and Schreiber (LMS)
\cite{Lev91} that the semi-inclusive
leptoproduction of
hadrons in deep inelastic electron or muon scattering can be used as
a check of SU(2) asymmetry in the proton sea. We will show that
the breakdown of p-n isospin symmetry could  also contribute to this
process and account for the data. According to the quark-parton model
considerations\cite{Close},
the number of hadrons h produced by a scattering off
the nucleon N in a given bin of
$x$ Bjorken variable and $z=E_{h}/\nu$ should be, up to a constant
factor:
\begin{equation}
N^{Nh}(x,z)\equiv \sum_{i}e^{2}_{i}q_{i}^{N}(x)D_{i}^{h}(z),
\label{eq:NNh}
\end{equation}
where $q_{i}^{N}(x)$ is the distribution function for quarks of
flavour i in the nucleon N and $D_{i}^{h}$ is the fragmentation
function of a quark with flavour i into the hadron h with energy
$E_{h}=z\nu$.
Defining
\begin{equation}
Q(z)=\frac{N^{p+}-N^{n+}+N^{p-}-N^{n-}}
{N^{p+}-N^{n+}-N^{p-}+N^{n-}},
\label{eq:LMS}
\end{equation}
where $N^{N\pm}=\int dx \,N^{N\pm}(x,z)$ is for
$\pm$ charged hadrons,
LMS found, assuming p-n isospin symmetry(i.e., Eq.(\ref{eq:pnis})),
\begin{equation}
Q(z)=3S_{G}\frac{4[D_{u}^{+}(z)+D_{\overline{u}}^{+}(z)]
                 -[D_{d}^{+}(z)+D_{\overline{d}}^{+}(z)]}
                {4[D_{u}^{+}(z)-D_{\overline{u}}^{+}(z)]
                 -[D_{d}^{+}(z)-D_{\overline{d}}^{+}(z)]},
\label{eq:Qz1}
\end{equation}
which is directly proportional to the outcome of the GSR
%Gottfried sum rule
in inclusive leptoproduction. After some detailed examination
of charged pions, kaons,
%and
protons and anti-protons, LMS
%Levelt, Mulders and Schreiber
derived an expression for $Q^{ch}(z)$,
using $D_{u}^{+}=D_{u}^{\pi+}+D_{u}^{K+}+D_{u}^{p}$,
\begin{equation}
zQ^{ch}(z)=3S_{G}z\frac{0.50z^{2}+3.1z+7.6}
{3.2z^{2}+11z+0.84}.
\label{eq:zQch1}
\end{equation}
They found that the data is consistent with both $S_{G}=0.24$ and
$S_{G}=1/3$.

We now examine the case that the violation of  GSR
%Gottfried sum rule violation
is due to isospin symmetry breaking between
%the sea quarks for
the proton and
%those for
the neutron while preserving  SU(2) symmetry in the sea
of nucleons.
%and the isospin symmetry between the valence quarks for protons and those
%for neutrons.
It can be seen, from Eq.(\ref{eq:NNh}),
\begin{equation}
\begin{array}{clcr}
N^{p\pm}=\frac{4}{9}u^{p}D_{u}^{\pm}
         +\frac{4}{9}\overline{u}^{p}D_{\overline{u}}^{\pm}
         +\frac{1}{9}d^{p}D_{d}^{\pm}
         +\frac{1}{9}\overline{d}^{p}D_{\overline{d}}^{\pm}
         +\frac{1}{9}s^{p}D_{s}^{\pm}
         +\frac{1}{9}\overline{s}^{p}D_{\overline{s}}^{\pm};\\
N^{n\pm}=\frac{4}{9}u^{n}D_{u}^{\pm}
         +\frac{4}{9}\overline{u}^{n}D_{\overline{u}}^{\pm}
         +\frac{1}{9}d^{n}D_{d}^{\pm}
         +\frac{1}{9}\overline{d}^{p}D_{\overline{d}}^{\pm}
         +\frac{1}{9}s^{n}D_{s}^{\pm}
         +\frac{1}{9}\overline{s}^{n}D_{\overline{s}}^{\pm}.
\end{array}
\end{equation}
Assuming charge conjugation invariance one has
$D_{u}^{\pm}=D_{\overline{u}}^{\mp}$ and
$D_{d}^{\pm}=D_{\overline{d}}^{\mp}$.
The flavour number conservation requires $u_{V}=u^{p}-\overline{u}^{p}
=d^{n}-\overline{d}^{n}=2$ and $d_{V}=d^{p}-\overline{d}^{p}
=u^{n}-\overline{u}^{n}=1$, thus, one obtains,
\begin{equation}
\begin{array}{clcr}
N^{p+}-N^{n+}=\frac{4}{9}D_{u}^{+}(1-\delta \overline{q})
         -\frac{4}{9}D_{\overline{u}}^{+}\delta \overline{q}
         -\frac{1}{9}D_{d}^{+}(1+\delta \overline{q})
         -\frac{1}{9}D_{\overline{d}}^{+}\delta \overline{q};\\
N^{p-}-N^{n-}=\frac{4}{9}D_{\overline{u}}^{+}(1-\delta \overline{q})
         -\frac{4}{9}D_{u}^{+}\delta \overline{q}
         -\frac{1}{9}D_{\overline{d}}^{+}(1+\delta \overline{q})
         -\frac{1}{9}D_{d}^{+}\delta \overline{q},
\end{array}
\end{equation}
from which we get,
\begin{equation}
Q(z)=\frac{4[D_{u}^{+}(z)+D_{\overline{u}}^{+}(z)](1-2\delta \overline{q})
       -[D_{d}^{+}(z)+D_{\overline{d}}^{+}(z)](1+2\delta \overline{q})}
                {4[D_{u}^{+}(z)-D_{\overline{u}}^{+}(z)]
                 -[D_{d}^{+}(z)-D_{\overline{d}}^{+}(z)]},
\label{eq:Qz2}
\end{equation}
where $\delta \overline{q}=\overline{q}^{n}-\overline{q}^{p}$ is the excess of
sea quarks in the neutron over those in the proton.
It can be seen that Eq.(\ref{eq:Qz2}) is different from
Eq.(\ref{eq:Qz1})
unless $S_{G}=1/3$ (i.e., $\overline{d}-\overline{u}=0$ in the SU(2)
sea asymmetry explanation and $\delta
\overline{q}=\overline{q}^{n}-\overline{q}^{p}=0$ in the p-n isospin symmetry
breaking explanation). Following the similar analyses of
Ref.\cite{Lev91} we find that the expressions $Q^{\pi}(z)$ and
$Q^{p}(z)$ are the same as for an asymmetric SU(2).
% sea asymmetry explanation.
However, $Q^{K}(z)$ is now
%found to be
\begin{equation}
Q^{K}(z)=\frac{(D^{K}+\tilde{D}^{K})(\frac{2}{5}+\frac{9}{5}S_{G})
-\frac{1}{2}\tilde{D}'^{K}(\frac{8}{5}-\frac{9}{5}S_{G})}
{D^{K}-\tilde{D}^{K}},
\end{equation}
which is different from Eq.(24) of Ref.\cite{Lev91}
unless $S_{G}=1/3$. Thus we
have, for $D_{u}^{+}=D_{u}^{\pi+}+D_{u}^{K+}+D_{u}^{p}$,
\begin{equation}
zQ^{ch}(z)=3S_{G}z\frac{0.80z^{2}+3.37z+7.63}
{3.2z^{2}+11z+0.84}
\label{eq:zQch2}
\end{equation}
for $S_{G}=0.24$ in the case of isospin symmetry breaking between
proton and neutron. Comparing Eq.(\ref{eq:zQch1}) and
Eq.(\ref{eq:zQch2})
we see that the expressions for $Q^{ch}(z)$ are different for the two
explanations.
In the later case the result is between
the GSR
%ottfried sum rule
prediction and the prediction for an
asymmetric SU(2) sea, and thus is also
consistent with the data of Ref.\cite{Lev91}.
To distinguish
between an asymmetric SU(2) sea  and p-n isospin symmetry breaking
from $Q^{ch}(z)$, we need
%more precise
data with better accuracy.
% are required.

\renewcommand{\thesection}{\Roman{section}.}
\section{SUMMARY}
\renewcommand{\thesection}{\arabic{section}}

We have examined several processes which have been suggested
to check whether the violation of GSR
%ottfried sum rule
is due to
an asymmetric SU(2) sea,
% asymmetry explanation of the Gottfried sum rule violation,
and found that
these processes are also sensitive to any excess of
sea quarks in the neutron over those in the proton.
% for the p-n isospin symmetry breaking explanation.
%, which is an alternative
%mechanism to explain the Gottfried sum rule violation.
%It has been found that the
A linear combination of neutrino structure functions
from protons and deuterons (i.e., Eq.(\ref{eq:KL})),
and the ratio $Q^{ch}(z)$
(i.e., Eq.(\ref{eq:LMS}))
for semi-inclusive leptoproduction of charged hadrons
can distinguish between the two possible explanations of the NMC
data, namely,
%have different values for the
an asymmetric SU(2) sea and p-n isospin symmetry, or
combinations of both.
%breaking explanation.
The values for the case of p-n isospin symmetry
breaking is between those for the GSR
%Gottfried sum rule
and those for the
SU(2) sea asymmetry explanation.
The quantities $R_{sea}$ (i.e., Eq.(\ref{eq:pidy})) in charged pion
Drell-Yan processes and $A_{DY}$ (i.e., Eq.(\ref{eq:ady}))
in proton Drell-Yan processes
give approximately the same values for the two different
explanations and thus can not settle the origin of the
GSR
%Gottfried sum rule
violation.
%It can be seen that the above processes are less sensitive
%as compared with the new sum(i.e., Eq.(\ref{eq:ns})) of the
The specific combination of neutrino and anti-neutrino scattering
data on protons and deuterons proposed in
Ref.\cite{Ma92} seems well suited to distinguish between the two
alternative explanations.
%flavour distribution
%asymmetry and isospin symmetry breaking explanations.
% of the   Gottfried sum rule violation.

%\noindent
%{\large \bf ACKNOWLEDGMENTS}

\break
\noindent
%{\large \bf Figure Captions}
%\renewcommand{\theenumi}{\ Fig.\arabic{enumi}}
%\begin{enumerate}
%\item
%\item
%\end{enumerate}

\end{document}